\documentclass[12pt]{article}
\newcommand{\lesssim}{\:\mbox{\raisebox{-3pt}{$\stackrel%
{\displaystyle <}{\sim}$}}\:}

\newcounter{saveeqn}
\newcommand{\alpheqn}%
    {\setcounter{saveeqn}{\value{equation}}%
     \stepcounter{saveeqn}%
     \setcounter{equation}{0}%
     \renewcommand{\theequation}{\arabic{saveeqn}\alph{equation}}}
\newcommand{\reseteqn}%
    {\setcounter{equation}{\value{saveeqn}}%
     \renewcommand{\theequation}{\arabic{equation}}}
\begin{document}

\title{\normalsize \hfill IMSc/2001/04/13\\
\normalsize \hfill UWThPh-2001-14\\
\normalsize \hfill IFIC/01-18\\[1cm] \LARGE
The solar LMA neutrino oscillation solution\\ in the Zee model}

\author{K.R.S. Balaji\footnote{E-mail: balaji@imsc.ernet.in}\\
\small Institute of Mathematical Sciences \\
\small Chennai 600\,113, India \\[3mm]
W. Grimus\footnote{E-mail: grimus@doppler.thp.univie.ac.at}\\ 
\small Universit\"at Wien, Institut f\"ur Theoretische Physik \\
\small Boltzmanngasse 5, A--1090 Wien, Austria\\[3mm]
T. Schwetz\footnote{E-mail: schwetz@flamenco.ific.uv.es}\\
\small Instituto de F\'{\i}sica Corpuscular - C.S.I.C., 
       Universitat de Val\`encia \\
\small Edificio Institutos de Paterna, Apt.\ 22085, E--46071 Val\`encia,
       Spain}

\date{3 April 2001}

\maketitle

\begin{abstract}
We examine the neutrino mass matrix in the version of Zee model where
both Higgs doublets couple to the leptons. We show that in this case
one can accommodate the large mixing angle (LMA) MSW solution of the solar
neutrino problem, while avoiding maximal solar mixing and
conflicts with constraints on lepton
family number-violating interactions. In the simplified scenario we
consider, we have the neutrino mass spectrum characterized by
$m_1 \simeq m_2 \simeq \sqrt{\Delta m^2_\mathrm{atm}}/\sin 2\theta$ and
$m_3/m_1 \simeq \cos 2\theta$, where $\theta$ is the solar mixing angle.
\end{abstract}

\newpage

\section{Introduction}

The results of the atmospheric \cite{SK} and solar neutrino experiments (for a
recent presentation of the solar results see, e.g., Ref.~\cite{GG}) indicate
that neutrinos have non-zero masses. A suitable environment for
obtaining Majorana neutrino masses is to extend the Higgs sector of the
Standard Model \cite{konetschny}. Extending the Standard Model with a singly
charged gauge singlet scalar and adding a second Higgs doublet allows to write
down an explicitly lepton number-violating interaction in the Higgs potential
and leads to 1-loop neutrino masses \cite{zee}. 
In the following we will only consider the lepton sector of the Zee model.

The Zee model is traded in two versions in the literature: the original model
\cite{zee}, which we will call general Zee model (GZM) in this paper, and a 
simpler version of the Zee model where only one of the two
Higgs doublets couples in the lepton sector \cite{wolfenstein}. The
latter version, which we will call restricted Zee model (RZM), can
naturally be achieved with a discrete symmetry \cite{grimus91}; it has the
advantage that the family lepton number-violating interactions mediated by the
couplings of the three physical neutral scalars are absent. The interesting
point is that the RZM leads to a symmetric neutrino mass matrix with zeros in
the diagonal\footnote{We always work in a basis where the charged lepton mass
matrix is diagonal.}; this mass matrix is called \emph{Zee mass matrix} in the
literature. One can easily check that all phases in the Zee mass matrix 
can be absorbed into the left-handed neutrino fields and, therefore, in
neutrino oscillations no CP violation is observable if the Zee mass matrix is
the correct neutrino mass matrix. 

The Zee mass matrix has a special feature: it has
been shown \cite{jarlskog,frampton} that it allows to accommodate only
bimaximal mixing, i.e., solar and atmospheric mixing angles are both very
close to $45^\circ$. While this is in excellent agreement with the atmospheric
neutrino data, results from the solar neutrino experiments are not that well
compatible with a solar mixing angle $\theta \simeq 45^\circ$, though this
value is also not excluded \cite{GG,GG-nir}. The purpose of the
present paper is to show that 
in the GZM it is no problem to accommodate the large mixing angle (LMA) MSW
solution of the solar neutrino deficit, while at the same time all
constraints on the additional couplings in the lepton sector stemming from
the second Higgs doublet are respected. In the GZM also the diagonal
elements of the neutrino mass matrix are non-zero in general.
Since the GZM is a quite rich and intricate model,
we restrict ourselves rather to an ``existence proof'' of the LMA MSW solution
within the GZM instead of discussing the GZM in full generality.

Let us outline our procedure:
\newcounter{point}
\begin{list}{\textit{Point \arabic{point}:}}{\usecounter{point}}
\item \label{Uassumptions}
We assume that $U_{e3} = 0$ ($U$ is the neutrino mixing matrix) and
the atmospheric mixing angle is exactly 
$45^\circ$ \cite{GG-valle}. Then in the mixing matrix the only free
parameter is the 
solar mixing angle $\theta$. This gives a certain form of the mass
matrix with 4 complex parameters (Section \ref{mm}).
\item \label{restricted}
The general problem is still quite intricate, so we set one of these
parameters of the mass matrix equal to zero and assume that the
remaining ones are real. In this scenario we can relate in a
simple way the three real parameters with the physical quantities
$\theta$, the solar mixing angle, $\Delta m^2_\odot = |m_1^2-m_2^2|$,
the solar mass-squared difference, 
$\Delta m^2_\mathrm{atm} = |m_3^2 - (m_1^2+m_2^2)/2|$, the atmospheric
mass-squared difference and $m_3$ ($m_j$ with $j=1,2,3$ are the
neutrino masses) and it is
possible to have the LMA angle solution of the solar neutrino
problem\footnote{This means that $\theta$ is large but safely 
smaller than $45^\circ$.} (Section \ref{simplemm}).
\item
Now the GZM is brought into play. After a discussion of the
neutrino mass matrix in this model (Section \ref{zeemm})
we assume that all quantities in the model
are real and we set to zero all but two of the additional Yukawa couplings
present in the GZM; this has the purpose of avoiding as many
family lepton number-violating neutral scalar interactions a possible. We show
that with these two additional 
coupling constants the restricted mass matrix mentioned in
Point~\ref{restricted}, which allows for the LMA MSW solution, can be
accommodated (Section \ref{zeemm}). 
\item
We complete our procedure by a numerical discussion of the parameters of our
scenario and by estimates of the rates of (family) lepton
number-violating processes (Section \ref{parameters}).
\end{list}
We also review the features of the RZM as a limit of the GZM 
(Section \ref{limit}) 
and summarize the results (Section \ref{summary}). In the appendix we
present the general formulas for the 1-loop Majorana neutrino mass
matrix induced by charged scalar loops.

\section{Neutrino mixing and the mass matrix}
\label{mm}

The Majorana neutrino mass matrix $\mathcal{M}_\nu$ is diagonalized
with a unitary matrix $V$ by
\begin{equation}\label{diag}
V^T \mathcal{M}_\nu V = \hat m = \mathrm{diag} \, (m_1,m_2,m_3) \,.
\end{equation}
With the assumptions mentioned in the introduction in Point
\ref{Uassumptions}, we can write the matrix $V$ as
\begin{equation}\label{U}
V = e^{i \hat \alpha} U e^{i \hat \beta}
\quad \mbox{with} \quad
U = \left(
\begin{array}{rrc}
\cos \theta & \sin \theta & 0 \\
-\sin \theta/\sqrt{2} & \cos \theta/\sqrt{2} & 1/\sqrt{2} \\
\sin \theta/\sqrt{2} & -\cos \theta/\sqrt{2} & 1/\sqrt{2} 
\end{array} \right) \,.
\end{equation}
The phases in $V$ suggest the definitions
\begin{equation}\label{mu}
\mathcal{M}'_\nu = U \hat \mu U^T
\quad \mbox{with} \quad 
\mathcal{M}'_\nu \equiv 
e^{i \hat \alpha} \mathcal{M}_\nu e^{i \hat \alpha} 
\quad \mbox{and} \quad
\hat \mu = e^{-2i\hat \beta} \hat m \,.
\end{equation}
Then $\mathcal{M}'_\nu$ can be expressed by $\hat \mu$ and the
parameters of $U$ in the following way:
\begin{equation}\label{mnu'}
\mathcal{M}'_\nu = \left(
\begin{array}{ccc} 
c^2 \mu_1 + s^2 \mu_2 & -cs(\mu_1 - \mu_2)/\sqrt{2} & 
cs(\mu_1 - \mu_2)/\sqrt{2} \\
-cs(\mu_1 - \mu_2)/\sqrt{2} & (s^2 \mu_1 + c^2 \mu_2 + \mu_3)/2 &
(-s^2 \mu_1 - c^2 \mu_2 + \mu_3)/2 \\
cs(\mu_1 - \mu_2)/\sqrt{2} & (-s^2 \mu_1 - c^2 \mu_2 + \mu_3)/2 &
(s^2 \mu_1 + c^2 \mu_2 + \mu_3)/2
\end{array}
\right)
\end{equation}
with $c \equiv \cos \theta$ and $s \equiv \sin \theta$ (see, e.g.,
Refs.~\cite{jarlskog,ma}). Thus, the mass matrix has the structure
\begin{equation}\label{M}
\mathcal{M}'_\nu = \left(
\begin{array}{rrr}
a & -b & b \\ -b & c & d \\ b & d & c
\end{array}
\right) \,.
\end{equation}
Consequently, for having $U_{e3} = 0$ and an atmospheric mixing angle
of exactly $45^\circ$ the following conditions on $\mathcal{M}'_\nu$
are necessary:
\alpheqn
\begin{eqnarray}
\mbox{Condition 1:} && \label{cond1}
\mathcal{M}'_{\nu\, e\mu} + \mathcal{M}'_{\nu\, e\tau} = 0 \,, \\
\mbox{Condition 2:} && \label{cond2}
\mathcal{M}'_{\nu\, \mu\mu} = \mathcal{M}'_{\nu\, \tau\tau} \,, \\
\mbox{Condition 3:} && \label{cond3}
\tan 2\theta = 2\sqrt{2} \frac{b}{a-c+d} \, \in \, \mathbf{R} \,.
\end{eqnarray}
\reseteqn
Finally, with the parameterization (\ref{M}) the complex masses
$\mu_j$ are found as
\alpheqn
\begin{eqnarray}
\mu_1 & = & \frac{1}{2} 
\left( a+c-d \pm \left[ (a-c+d)^2 + 8b^2 \right]^{1/2} \right) \,,
\label{mu1} \\
\mu_2 & = & \frac{1}{2} 
\left( a+c-d \mp \left[ (a-c+d)^2 + 8b^2 \right]^{1/2} \right) \,,
\label{mu2} \\
\mu_3 & = & c+d \,. \label{mu3}
\end{eqnarray}
\reseteqn

\section{A simplified mass matrix}
\label{simplemm}

Having discussed the general form of the mass matrix which leads to
the mixing matrix (\ref{U}), we now investigate the consequences of the
following simplifying assumptions:
\begin{equation}\label{simple}
a, b, d \in \mathbf{R} \quad \mbox{and} \quad c=0 \,.
\end{equation}
In the next section, this scenario will be reproduced in the framework
of the Zee model. With the reality assumptions, the quantities $\mu_j$
(\ref{mu}) are identical with the neutrinos masses apart from possible
signs. The experimentally accessible quantities are expressed as
\alpheqn
\begin{eqnarray}
\tan 2\theta & = & 2\sqrt{2} \frac{b}{a+d} \label{theta} \,, \\
\Delta m^2_\odot & = & |a-d| \left[ (a+d)^2 + 8b^2 \right]^{1/2} \,,
\label{dsolar} \\
\Delta m^2_\mathrm{atm} & = & \frac{1}{2} (a^2-d^2) + 2b^2 \,,
\label{datm} \\
m_3 & = & |d| \label{m3d}
\end{eqnarray}
\reseteqn
by the parameters $a$, $b$, $d$. We have chosen $m_3$ as representative
of the absolute neutrino mass values, since it is simply given by
Eq.~(\ref{m3d}). Without loss of generality we
will adopt henceforth the following conventions: 
$0^\circ \leq \theta \leq 90^\circ$, $m_1 < m_2$, $\mu_3 = m_3$. 
It follows from the last relation and from Eq.~(\ref{mu3}) that $d$ is
positive. Note that in Eq.~(\ref{datm}) no absolute value of
the right-hand side of the equation is necessary, because it must be
positive. The argument goes as follows. Suppose that
$(a^2-d^2)/2 + 2b^2 = -\Delta m^2_\mathrm{atm}$. Then it follows that
$d^2-a^2 \geq 2\Delta m^2_\mathrm{atm}$. Therefore, $d^2-a^2$ is
positive, which allows to derive the inequality 
$\Delta m^2_\odot \geq d^2-a^2 \geq 2\Delta m^2_\mathrm{atm}$ from
Eq.~(\ref{dsolar}). This is a contradiction to the values of the
mass-squared differences, fitted from the data \cite{GG}.

In Eqs.~(\ref{theta}), (\ref{dsolar}), (\ref{datm}) and (\ref{m3d}),
four physical quantities are expressed by three parameters. Therefore,
a consistency condition exists, which is given by
\begin{equation}\label{C}
\Delta m^2_\mathrm{atm} = 
\frac{1}{2} \eta \Delta m^2_\odot |\cos 2\theta| + \frac{1}{4}
\left( \sqrt{m_3^2 + \eta \Delta m^2_\odot |\cos 2\theta|} +
\eta' m_3 \right)^2 \tan^2 2\theta \,.
\end{equation}
The signs $\eta$ and $\eta'$ occurring in this equation are 
$\eta = \mbox{sign}\, (a^2-d^2)$ and $\eta' = \mbox{sign}\, a$.
In the context of the Zee model we will finally need the relations
\alpheqn
\begin{eqnarray}
a^2 & = & m_3^2 + \eta \Delta m^2_\odot |\cos 2\theta| \,, \label{a} \\
b^2 & = & \frac{1}{2} \Delta m^2_\mathrm{atm} - \frac{1}{4} \eta
\Delta m^2_\odot |\cos 2\theta| \,. \label{b}
\end{eqnarray}
\reseteqn

Looking at the consistency condition (\ref{C}) and assuming that
$m_3^2$ is of the order of $\Delta m^2_\odot |\cos 2\theta|$ or smaller,
we obtain 
$\sin^2 2\theta/|\cos 2\theta| \sim \Delta m^2_\mathrm{atm}/\Delta
m^2_\odot$, which 
amounts to bimaximal mixing. Since we want to show that the Zee model
allows to avoid bimaximal mixing we concentrate on
\begin{equation}\label{basic}
m_3^2 \gg \Delta m^2_\odot |\cos 2\theta| \,.
\end{equation}
With this assumption it is easy to obtain an approximate expression for
$m_3^2$. One can check that for $\eta' = -1$ one arrives again at bimaximal
mixing. Using $\eta' = 1$, we have $a > 0$ and we easily calculate
\begin{equation}\label{m3}
m_3^2 = \Delta m^2_\mathrm{atm} \left\{
\frac{1}{\tan^2 2\theta} - 
\eta \frac{1}{2} \frac{\Delta m^2_\odot}{\Delta m^2_\mathrm{atm}} 
\frac{|\cos 2\theta|}{\sin^2 2\theta} + \frac{1}{16} 
\left( \frac{\Delta m^2_\odot}{\Delta m^2_\mathrm{atm}} \right)^2
\sin^2 2\theta + \ldots \right\} \,.
\end{equation}
Using this equation the condition (\ref{basic}) implies
\begin{equation}\label{basic1}
|\cos 2\theta| \gg \frac{\Delta m^2_\odot}{\Delta m^2_\mathrm{atm}} \,.
\end{equation}
Equation (\ref{m3}) together with $a > 0$ and $d > 0$,
allows to estimate from Eq.~(\ref{a})
that 
\begin{equation}\label{a-d}
a-d \simeq \frac{1}{2} \eta \frac{\Delta m^2_\odot}{\sqrt{\Delta
m^2_\mathrm{atm}}} \sin 2\theta \,.
\end{equation}
From Eqs.~(\ref{theta}) and (\ref{dsolar}) and the convention 
$m_1 < m_2$, we can express the masses $m_1$ and $m_2$ as
\alpheqn
\begin{eqnarray}
m_1 & = & \frac{1}{2} \left( -|a-d| + \frac{\Delta m^2_\odot}{|a-d|}
\right) \,, \label{m1} \\
m_2 & = & \frac{1}{2} \left( \hphantom{-} |a-d| + 
\frac{\Delta m^2_\odot}{|a-d|} \right) \,. \label{m2}
\end{eqnarray}
\reseteqn
Inspection of Eqs.~(\ref{mu1}) and (\ref{mu2}) reveals that our
conventions fix the signs of $\mu_{1,2}$:
$\mbox{sign}\, \mu_1 = -\mbox{sign}\, \mu_2 = -\eta$.
Then, with Eq.~(\ref{a-d}), an estimate of the neutrino masses which neglects
the solar mass-squared difference is given by
\begin{equation}\label{estimate}
m_1 \simeq m_2 \simeq \frac{\sqrt{\Delta m^2_\mathrm{atm}}}{\sin 2\theta} 
\quad \mbox{and} \quad 
m_3 \simeq \frac{\sqrt{\Delta m^2_\mathrm{atm}}}{|\tan 2\theta|} \,.
\end{equation}
This equation tells us that $m_3 < m_{1,2}$, at least in the regime of large
solar mixing.

\section{Neutrino masses in the general Zee model}
\label{zeemm}

In the previous section we have discussed the mass matrix determined by
Eqs.~(\ref{M}) and (\ref{simple}) without reference to any specific model of
neutrino masses. Now we introduce the Zee model \cite{zee} and discuss the
neutrino mass matrix in the case that both scalar doublets of the Zee model
couple in the lepton section. The Yukawa Lagrangian is given by
\begin{equation}
\mathcal{L}_Y = -\sum_{a=1}^2 \bar L \Gamma_a \phi_a \ell_R +
L^T C^{-1} i\tau_2 f L h^+ + \mathrm{h.c.},
\end{equation}
where $f$ is an antisymmetric $3 \times 3$ matrix \cite{zee}. The mass matrix
of the charged leptons arises at tree level through
\begin{equation}
\langle \phi_a \rangle_0 = \frac{v_a}{\sqrt{2}}
\left( \begin{array}{c} 0 \\ 1 \end{array} \right)
\quad \mbox{and} \quad
M_\ell = \frac{1}{\sqrt{2}} \sum_{a=1}^2 v_a \Gamma_a 
\end{equation}
with $v \equiv \sqrt{|v_1|^2 + |v_2|^2} \simeq 246$ GeV.
The physical charged scalar fields $H^+_1$, $H^+_2$ with masses $M_1$,
$M_2$, respectively, and the would-be Goldstone
boson $\phi_w^+$ are obtained by the unitary transformation
\cite{zee,grimus91,grimus90} 
\begin{equation}
\left( \begin{array}{c} \phi^+_1 \\ \phi^+_2 \\ h^+ \end{array}
\right) =
\left( 
\begin{array}{ccc} 
\frac{v_1}{v} & -\frac{v_2^*}{v} W_{11} & -\frac{v_2^*}{v} W_{12} \\
\frac{v_2}{v} & \frac{v_1^*}{v} W_{11} & \frac{v_1^*}{v} W_{12} \\
0 & W_{21} & W_{22}
\end{array}
\right) \left(
\begin{array}{c} \phi_w^+ \\ H^+_1 \\ H^+_2 \end{array} \right).
\end{equation}
As anticipated in the introduction, we assume to be in a basis where
the charged lepton mass matrix is diagonal, i.e., $M_\ell = \hat{M}_\ell$,
where the hat symbolizes that this mass matrix is diagonal. In this
basis we have 
\begin{equation}
\Gamma_1 = \frac{1}{v_1} ( \sqrt{2} \hat{M}_\ell - v_2 \Gamma_2)\,.
\end{equation}
In our parameterization the 
flavour-changing Higgs couplings are given by the off-diagonal
elements of $\Gamma_2$. Furthermore, off-diagonal elements of the
2$\times$2 unitary matrix $W$ are present because of the vacuum
expectation value of the lepton number-violating term 
\begin{equation}\label{potential}
\lambda h^+ \phi_1^\dagger \tilde{\phi}_2 + \mbox{h.c.} 
\end{equation}
in the Higgs potential.

The physical charged Higgses couple to the leptons in the following way:
\begin{eqnarray}
\lefteqn{-\mathcal{L}_Y(H^+) =} \nonumber \\
&& \bar\nu_L \left( -\sqrt{2} \frac{v_2^*}{v_1v} \hat M_\ell +
\frac{v}{v_1} \Gamma_2 \right) 
\ell_R (W_{11} H_1^+ + W_{12} H_2^+ ) \nonumber \\
&& + \overline{(\nu_L)^c} (2f) \ell_L (W_{21} H_1^+ + W_{22} H_2^+ ) 
+ \mbox{h.c.}
\label{H+coupling}
\end{eqnarray}
With the formulas in the appendix we obtain \cite{zee}
\begin{equation}\label{Mnuzee}
\mathcal{M}_\nu = \sum_{j=1}^2 A_L^j \hat{M}_\ell
I(M_j^2,\hat{M}_\ell^2) {A_R^j}^\dagger + \mbox{transp.}
\end{equation}
with
\begin{equation}
A_R^j = \left( -\sqrt{2} \frac{v_2^*}{v_1v} \hat M_\ell +
\frac{v}{v_1} \Gamma_2 \right) W_{1j}
\quad \mbox{and} \quad A_L^j = 2f W_{2j}\,.
\end{equation}
The infinity in Eq.~(\ref{Mnuzee}) cancels \cite{zee} because
$\sum_{j=1}^2 W_{2j} W_{1j}^* = 0$. Defining
\begin{equation}
J(M_1^2, M_2^2, m^2) = 
\frac{M_1^2}{M_1^2 - m^2} \ln \frac{M_1^2}{m^2} -
\frac{M_2^2}{M_2^2 - m^2} \ln \frac{M_2^2}{m^2},
\end{equation}
from Eq.~(\ref{Mnuzee}) 
we obtain the final result \cite{zee,grimus90}
\begin{eqnarray}
\lefteqn{\mathcal{M}_\nu = 2 W_{21} W_{11}^* \times \frac{1}{(4\pi)^2}} 
\nonumber \\
&& \times
\left\{ \sqrt{2} \frac{v_2}{v_1^* v} 
\left( \hat{M}_\ell^2 \hat{J} f - f \hat{J} \hat{M}_\ell^2 \right) -
\frac{v}{v_1^*} \left( \gamma \hat{M}_\ell \hat{J} f -
f \hat{J} \hat{M}_\ell \gamma^T \right) \right\}
\label{final}
\end{eqnarray}
with $\hat J \equiv \mbox{diag}\, (J(M_1^2,M_2^2,m_\alpha^2))$ 
($\alpha = e, \mu, \tau$) and $\gamma \equiv \Gamma_2^*$.
Note that for $m^2 \ll M^2_{1,2}$ the function $J$ simplifies to 
\begin{equation}\label{Japprox}
J \simeq 2\ln (M_1/M_2) \,.
\end{equation}
For the product $W_{21} W_{11}^*$ of elements of the charged-scalar
mixing matrix $W$, we obtain the relation
\begin{equation}\label{WW}
W_{21}W_{11}^* = \frac{\lambda^* v}{\sqrt{2}(M_1^2-M_2^2)} \,.
\end{equation}
It shows explicitly that the Majorana neutrino masses are proportional to
the coupling $\lambda$ in the Higgs potential (\ref{potential}).

In the GZM considered here, there are family
lepton number-violating processes induced by the charged and the
neutral scalar interactions. Experimental bounds constrain the
coupling matrices $f$ and $\Gamma_2$. 

\section{The simplified mass matrix within the Zee model}

In this section, our aim is to reproduce the neutrino mass matrix defined by
Eqs.~(\ref{M}) and (\ref{simple}) within the GZM. In order to save the
amount of writing we introduce the notation
\begin{equation}\label{MAB}
\mathcal{M}_\nu = A \left( \hat{r}^2 f - f \hat{r}^2 \right) -
B \left( \gamma \hat{r} f - f \hat{r} \gamma^T \right)
\quad \mbox{with} \quad
\hat{r} = \mbox{diag}\, (m_e, m_\mu, m_\tau)/v \,.
\end{equation} 
Both constants $A$ and $B$ are of the order of 1 GeV, resulting from
dividing the electroweak scale by $16 \pi^2$:
\begin{equation}\label{AB}
B = 2 W_{21} W_{11}^* \times \frac{1}{(4\pi)^2} \ln \frac{M_1^2}{M_2^2}
\times \frac{v^2}{v_1^*} 
\quad \mbox{and} \quad
A = \sqrt{2}\, \frac{v_2}{v} B \,.
\end{equation}
Since the off-diagonal elements of $\gamma$ introduce flavour-changing neutral
interactions, we adopt the philosophy to set to zero as many of them as
possible. As we will see, it turns out that
\begin{equation}\label{nonzero}
\gamma_{e\tau} \neq 0 \quad \mbox{and} \quad \gamma_{\tau\tau} \neq 0 \,,
\end{equation}
and all other elements of $\gamma$ being equal to zero, is sufficiently
general to avoid bimaximal mixing, which necessarily happens for $\gamma = 0$
\cite{jarlskog}. It can easily be checked that with this assumption
Condition 2 (\ref{cond2}) is fulfilled by having 
$\mathcal{M}'_{\nu\, \mu\mu} = \mathcal{M}'_{\nu\, \tau\tau} = 0$.
Furthermore, we assume that all quantities we deal
with are real: $A$, $B$, the elements of the matrix $f$, and
$\gamma_{e\tau}$ and $\gamma_{\tau\tau}$. Thus we identify
$\mathcal{M}_\nu$ with $\mathcal{M}'_\nu$.

We have to reproduce with the mass matrix (\ref{MAB}) the parameters
$a$ (\ref{a}), $b$ (\ref{b}), $d$ (\ref{m3d}) and Condition 1
(\ref{cond1}). Thus we have five coupling constants, three in $f$ and
$\gamma_{e\tau}$, $\gamma_{\tau\tau}$ and four relations. It is
convenient to express the other four coupling constants as a function
of $\gamma_{\tau\tau}$:
\alpheqn
\begin{eqnarray}
f_{e\mu} & = & \frac{b}{A(r_\mu^2-r_e^2)}
\left\{ 1 + \frac{ad}{2b^2}\, 
\frac{A (r_\tau^2-r_e^2  ) - B r_\tau \gamma_{\tau\tau}}%
     {A (r_\tau^2-r_\mu^2) - B r_\tau \gamma_{\tau\tau}}  \right\} \,,
\label{femu}  \\
f_{e\tau} & = & \frac{-b}%
                     {A (r_\tau^2-r_e^2) - B r_\tau \gamma_{\tau\tau}} \,,
\label{fetau} \\
f_{\mu\tau} & = & \frac{-d}%
                      {A (r_\tau^2-r_\mu^2) - B r_\tau
                      \gamma_{\tau\tau}} \,,
\label{fmutau} \\
\gamma_{e\tau} & = & -\frac{a}{2B r_\tau b}
\left\{ A (r_\tau^2-r_e^2) - B r_\tau \gamma_{\tau\tau} \right\} \,.
\label{getau}
\end{eqnarray}
\reseteqn

\section{Numerical estimates}
\label{parameters}

Let us now estimate the values of the coupling constants. For
definiteness we take 
$\Delta m^2_\mathrm{atm} = 3 \times 10^{-3}$ eV$^2$ and $A = 2$ GeV.
Furthermore, we need the values of 
$r_\tau^2 \simeq 5.22 \times 10^{-5}$ and 
$m_\mu^2/m_\tau^2 \simeq 3.535 \times 10^{-3}$.
Defining $x = B \gamma_{\tau\tau}/A r_\tau$ and assuming
that\footnote{This avoids some finetuning for
$\gamma_{\tau\tau}$. Note, however, that $x=0$ is also possible.}
$x \sim 1$ and
$(1-m_e^2/m_\tau^2-x)/(1-m_\mu^2/m_\tau^2-x) \simeq 1$, we obtain
\alpheqn
\begin{eqnarray}
f_{e\mu} & \simeq & \hphantom{-} 1.0 \times 10^{-4} \times 
\frac{\mbox{sign}\, b}{\sin^2 2\theta} \,,
\label{femu-num} \\
f_{e\tau} & \simeq & - 3.7 \times 10^{-7} \times
\frac{\mbox{sign}\, b}{1-x} \,, \label{fetau-num} \\
f_{\mu\tau} & \simeq & - 5.2\times 10^{-7} \times
\frac{1}{|\tan 2\theta|\, (1-x)} \,, \label{fmutau-num} \\
\gamma_{e\tau} & \simeq & - 7.2 \times 10^{-3} \times
\frac{v_2}{v}\, \frac{1-x}{\tan 2\theta} \,.
\end{eqnarray}
\reseteqn
These equations serve to see the orders of magnitude and any effects of
$\Delta m^2_\odot$ are neglected. As can be seen from Eq.~(\ref{a-d}),
a considerable amount of finetuning is involved in
order to reproduce the solar mass-squared difference. 

Now we concentrate on the LMA MSW solution of the solar
neutrino problem, where $\theta$ is in the first octant.
With Eq.~(\ref{theta}) it follows that $b>0$.
In this case a representative value of the mixing
angle $\theta$ is given by the best fit value $\sin^2 2\theta \simeq 0.75$
($\tan 2\theta \simeq 1.71$) of Ref.~\cite{GG}, with the corresponding
mass-squared difference 
$\Delta m^2_\odot \simeq 3.2 \times 10^{-5}$ eV$^2$.
We note that in this case we have
$|f_{e\tau}/f_{e\mu}| \sim m_\mu^2/m_\tau^2$, which is similar to the
case of the Zee mass matrix \cite{jarlskog}. On the other hand, in our
scenario we have $f_{e\tau} \sim f_{\mu\tau}$, whereas for the Zee
mass matrix the relation $|f_{e\tau}| \gg |f_{\mu\tau}|$ holds
\cite{jarlskog}. 
As far as $\gamma$ is concerned, with $x \sim 1$ we can have 
$\gamma_{e\tau} \sim \gamma_{\tau\tau} \sim 10^{-3} \div 10^{-2}$.

Due to our assumption (\ref{nonzero}),
flavour-changing neutral scalar interactions at the tree level are
very constrained. Among the charged lepton decays we only have
$\tau^- \to e^- \ell^+ \ell^-$ with $\ell = e, \mu$. A generous
estimate of the branching ratio of this decay for $\ell = \mu$ is
obtained by
\begin{equation}\label{br}
\mbox{Br}(\tau^- \to e^- \mu^+ \mu^-) \sim
\frac{1}{G_F^2 M_0^4} \gamma_{e\tau}^2 \left( \frac{m_\mu}{v}
\right)^2 < 10^{-8} \,,
\end{equation}
where we have taken $G_F^2 M_0^4 \sim 10^{-2}$ ($M_0 \sim 100$ GeV)
and $M_0$ is a generic neutral
Higgs mass. At the 1-loop level, neutral Higgs exchange also induces
the decay $\tau^- \to e^- \gamma$. Making again an estimate, we obtain
\cite{petcov,sher} 
\begin{equation}\label{br-rad}
\mbox{Br}(\tau^- \to e^- \gamma) \sim \frac{\alpha}{48\pi}\,
\frac{1}{G_F^2 M_0^4} (\gamma_{e\tau} \gamma_{\tau\tau})^2 <
10^{-10} \,.
\end{equation}
Both estimates are well compatible with the experimental upper bounds on these
branching ratios of the order of $10^{-6}$ \cite{groom}.
More detailed discussions of these decays are found in
Refs.~\cite{sher,kang}. We have used $|\gamma_{e\tau}| \lesssim 10^{-2}$
and $|\gamma_{\tau\tau}| \lesssim 10^{-2}$ in Eqs.~(\ref{br}) and
(\ref{br-rad}). 

Also the charged scalars participate in various charged
lepton decays as intermediate particles. Numerous decays of the type
$\ell^-_1 \to \ell^-_2 + 2\; \mbox{neutrinos}$ proceed at tree level via the
Lagrangian (\ref{H+coupling}). According to the couplings in this Lagrangian
we can distinguish between $f$, $\gamma$ and $\hat r$ vertices, and we can
have decay amplitudes with all possible combinations of these vertices, except
$\gamma$-$\gamma$, which is forbidden due to the restricted form of $\gamma$,
Eq.~(\ref{nonzero}). E.g., to the Standard Model amplitude of ordinary muon
decay there is an amplitude with couplings $r_\mu r_e$ and another one
with $f_{e\tau}f_{\mu\tau}$; another example is 
$\tau^- \to \bar\nu_e \bar\nu_\mu \mu^-$ with couplings $f_{e\tau} r_\mu$,
which is lepton number violating. The branching ratios of all these decays are
negligible because of the smallness of the coupling constants
$f_{\alpha\beta}$ and $\gamma_{e\tau}$ and the ratios $r_\alpha$. 
Also negligible are radiative decays 
$\ell_1^- \to \ell_2^- + \gamma$ induced by charged Higgs loops
\cite{petcov}. In this case one 
always has two $f$-vertices in the loop graph, except in the case of
the amplitude for $\tau^- \to e^- + \gamma$, where there are two
contributions, proportional to $f_{e\mu} f_{\mu\tau}$ and $\gamma_{e\tau} r_e$.
Recent reviews of the restrictions on the coupling constants
$f_{\alpha\beta}$ are found in Refs.~\cite{smirnov,mituda}.

Scalar contributions to the anomalous magnetic moments of the electron and
muon involving $f$ \cite{mituda} and $\hat r$-couplings are totally
negligible because these constants are too small. 
A contribution from the $\gamma$-couplings to the
electron magnetic moment coming from $\tau$ exchange is proportional to
$\gamma_{e\tau}\gamma_{\tau e}$ \cite{leveille} and is thus zero in view of
$\gamma_{\tau e} = 0$.

In our scenario we have $\mathcal{M}_{\nu\, ee} = a \neq 0$. 
The matrix element $\mathcal{M}_{\nu\, ee}$ is identical with the
effective neutrino mass $\langle m_\nu \rangle$ probed in neutrinoless
double-beta decay. Therefore, this decay is allowed and the
effective neutrino mass is given by
\begin{equation}
| \langle m_\nu \rangle | \simeq m_3 \simeq 
\frac{0.05\; \mbox{eV}}{\tan 2\theta} \,.
\end{equation}
This represents an order of magnitude which is accessible in future
experiments (for recent reviews see, e.g., Ref.~\cite{fiorini}).

Having seen that the LMA MSW solution of the solar neutrino problem can be
accommodated in our scenario, we now proceed to the small mixing angle (SMA)
MSW solution. In this case 
we take as illustration the best fit value of Ref.~\cite{GG}, 
$\sin^2 2\theta \simeq 2.3 \times 10^{-3}$ ($\tan 2\theta \simeq 4.8
\times 10^{-2}$), which has a corresponding $\Delta m^2_\odot \simeq
0.5 \times 10^{-5}$ eV$^2$. From Eq.~(\ref{femu-num}) we see that now
$f_{e\mu} \simeq 0.05$ becomes relatively large and barely compatible with
the requirement that the Zee boson does not have an effect on the muon
decay rate so that it does not destroy the agreement in electroweak
precision tests \cite{smirnov}.
Thus in our simple scenario we cannot incorporate safely
the SMA solution.

\section{The limit $\Gamma_2 \to 0$}
\label{limit}

To make contact with Refs.~\cite{jarlskog,frampton}, we explore the
effect of $\Gamma_2 = 0$. This means that all diagonal elements of the
neutrino mass matrix are zero \cite{zee} and, in our notation,
we have $a=0$, i.e., $m_3^2 = \Delta m^2_\odot |\cos 2\theta|$ 
and $\eta = -1$ (see (Eq.~(\ref{a})). 
These relations and inspection of the consistency
condition (\ref{C}) leads to
\begin{equation}\label{result1}
m_3 \simeq \frac{1}{2} \,
\frac{\Delta m^2_\odot}{\sqrt{\Delta m^2_\mathrm{atm}}} 
\,, \quad
m_1 \simeq m_2 \simeq \sqrt{\Delta m^2_\mathrm{atm}} 
\quad \mbox{and} \quad
|\cos 2\theta| \simeq \frac{1}{4} \,
\frac{\Delta m^2_\odot}{\Delta m^2_\mathrm{atm}} \,.
\end{equation}
From the last relation we read off that $\theta$ is 45$^\circ$ for all
practical purposes.
The consequences for the coupling matrix $f$ are
\begin{equation}\label{result2}
\left| \frac{f_{e\tau}}{f_{e\mu}} \right| \simeq 
\frac{m_\mu^2}{m_\tau^2}
\quad \mbox{and} \quad
\left| \frac{f_{e\tau}}{f_{\mu\tau}} \right| \simeq
\sqrt{2} \, \frac{\Delta m^2_\mathrm{atm}}{\Delta m^2_\odot} \,.
\end{equation}
The first of these two relations is obtained by taking the ratio of
Eq.~(\ref{fetau}) and Eq.~(\ref{femu}), where $a=0$ is taken into account. In
the second relation, Eq.~(\ref{b}) has been used.
The results (\ref{result1}) and (\ref{result2}) agree with those of
Refs.~\cite{jarlskog,frampton}. 

\section{Summary}
\label{summary}

In this paper we have discussed neutrino masses and mixing in the
general Zee model, where both Higgs doublets couple in the lepton
sector. In this endeavour, we were motivated by the result 
that for $\Gamma_2 = 0$, where $\Gamma_2$ denotes the Yukawa
coupling matrix of the second Higgs doublet, the
Zee mass matrix leads to bimaximal mixing. It is true that a mixing angle of
$45^\circ$ is perfect for the description of the atmospheric neutrino
data, but it does not represent a very good fit for the solar neutrino data.
The general Zee model is a rather rich and intricate model. Therefore,
in order to simplify the analysis, we have assumed that all
quantities appearing in the neutrino mass matrix are real and that in
the second Yukawa coupling matrix $\Gamma_2$, which is set to zero
usually, only two elements, the $e\tau$ and $\tau\tau$ elements, are
non-zero. We have shown that this is sufficient to accommodate 
solutions of the solar neutrino problem with a large mixing angle instead of
maximal mixing; on the other hand, in the atmospheric sector maximal
mixing remains. At the same time, all 
dangerous processes induced by the new couplings are sufficiently
suppressed. Actually, we could have even set the $\tau\tau$ element of
$\Gamma_2$ equal to zero and used a single non-zero element in this
coupling matrix. However, the $\tau\tau$ element represents a
possibility to achieve a further suppression of the potentially more
dangerous $e\tau$ element. As in the case $\Gamma_2 = 0$, the neutrino
mass $m_3$ is the smallest neutrino mass, and the resulting mass
spectrum is rather of the type which is called ``inverted
hierarchy''. However, whereas for $\Gamma_2 = 0$ one has $m_3 \ll m_1
\simeq m_2$, now 
$m_3 \simeq \sqrt{\Delta m^2_\mathrm{atm}}/\tan 2\theta$ is of the
same order of magnitude as the other two masses.

Numerous finetunings are involved in our scenario.
Implementing the condition (\ref{cond1}) and setting all but two elements
of $\Gamma_2$ equal to zero represent rather severe finetunings. 
These procedures were useful for exactly having $U_{e3}=0$ and
an atmospheric mixing angle of $45^\circ$, and for avoiding a
dangerous class of lepton family number-violating interactions.
In order to reproduce the neutrino masses within our scenario, we need
$|f_{e\tau}| \sim |f_{\mu\tau}| \sim |f_{e\mu}| (m_\mu/m_\tau)^2$ 
and a quite drastic finetuning between $a$ and $d$ in the mass matrix
(\ref{M}), in order to accommodate the solar mass-squared
difference. Of course, the mass-squared difference 
$\Delta m^2_\odot \sim 10^{-8}$ eV$^2$ of 
the quasi-vacuum oscillation solution \cite{GG} would require a stronger
finetuning than $\Delta m^2_\odot \sim 10^{-5}$ eV$^2$ of the large
mixing angle solution. Examination of the Zee model
with a general $\Gamma_2$ and with small deviations of the
neutrino mass matrix from the form (\ref{M}) should reveal how much
our finetunings could be relaxed. In the present work we have confined
ourselves to prove that it is possible to reproduce the LMA MSW
solution within the Zee model.

In conclusion, even in the case that bimaximal mixing is ruled
out, the Zee model will remain a viable and interesting scenario in
order to accommodate the neutrino oscillation solutions of the solar
and atmospheric neutrino problems.

\section*{Acknowledgements}

This work was supported by Spanish DGICYT under grant PB98-0693, by the 
European Commission TMR network HPRN-CT-2000-00148 and the ESF network 86. 
T.S.\ is supported by the Marie Curie Training Site Program for 
Particle Physics Beyond the Standard Model, contract number HPMT-2000-00124.

\setcounter{section}{0}
\setcounter{equation}{0}
\renewcommand{\theequation}{\Alph{section}\arabic{equation}}
\begin{appendix}
\section{Charged scalar exchange and the 1-loop neutrino mass}

We proceed from the general Lagrangian
\begin{equation}\label{Lphi}
-\mathcal{L}_Y(\nu_L,\ell,\varphi^+) = 
(\bar \nu_L A_R \ell_R + \overline{(\nu_L)^c} A_L \ell_L ) \varphi^+ 
+ \mbox{h.c.},
\end{equation}
where the $A_{L,R}$ are 3$\times$3 coupling matrices in the case of 3
families and $\ell_{L,R}$ denote chiral charged fermion fields. The
neutrino Majorana 
mass Lagrangian is defined by
\begin{equation}
\mathcal{L}_{\nu\, \mathrm{mass}} = 
\frac{1}{2} \nu_L^T C^{-1} \mathcal{M}_\nu \nu_L + \mbox{h.c.}
\end{equation}
With the scalar interactions (\ref{Lphi}) we obtain
\begin{equation}
\mathcal{M}_\nu^\mathrm{1-loop} = A_L \hat{M}_f
I(M_\varphi^2,\hat{M}_f^2) A_R^\dagger + \mbox{transp.},
\end{equation}
where we have defined
\begin{eqnarray}
\lefteqn{I(M^2,m^2) = \frac{1}{(4\pi)^2}} \nonumber \\ 
&& \times \left\{ - \frac{1}{\epsilon} + 
\ln (4\pi)  + \gamma_E -1 + \frac{1}{M^2 - m^2} 
\left( M^2 \ln \frac{M^2}{\mu^2} - m^2 \ln \frac{m^2}{\mu^2} \right)
\right\} \,. \nonumber \\ &&
\end{eqnarray}
We have used dimensional regularization; thus, 
$\epsilon = (4-n)/2$ with $n$ being the number of space-time dimensions,
$\gamma_E$ is Euler's constant and $\mu$
is an arbitrary mass scale.
\end{appendix}

\end{document}